\documentclass[]{article}
\usepackage[english]{babel}
\usepackage{epsfig,dsfont,multicol}
\usepackage{amsfonts,amssymb,amsmath,amsthm,scalefnt,mathrsfs}
\usepackage{graphicx,color}

\newcommand\barQ{{\bar Q}}
\newcommand\barnu{{\bar\nu}}

\newcommand\ml{\left\{\begin{array}}
\newcommand\mr{\end{array}\right\}}

\begin{document}

\title{A model capturing novel strand symmetries in bacterial DNA}
\author{Marcelo Sobottka$^a$\footnote{\noindent Corresponding author.\newline
 E-mail addresses: sobottka@mtm.ufsc.br (M. Sobottka), ahart@dim.uchile.cl (A. G. Hart).}\ , Andrew G. Hart $^b$\\
\footnotesize{$^a$ Departamento de Matem\'{a}tica, Universidade Federal de Santa Catarina, Brazil}\\
\footnotesize{$^b$ Departamento de Ingenier\'{\i}a Matem\'{a}tica and Centro de Modelamiento Matem\'{a}tico, Universidad de Chile, Chile.}\\
}

\date{\ }
\maketitle

\begin{abstract}
Chargaff's second parity rule for short oligonucleotides states that the
frequency of any short nucleotide sequence on a strand is
approximately equal to the frequency of its reverse complement on the same strand. Recent studies have shown
that, with the exception of organellar DNA, this parity rule generally holds for double
stranded DNA genomes and fails to hold for single-stranded genomes. While
Chargaff's first parity rule is fully explained by the Watson-Crick pairing in
the DNA double helix, a definitive explanation for the second parity rule has
not yet been determined. In this work, we propose a model based on
a hidden Markov process for approximating the distributional
structure of primitive DNA sequences. Then, we use the model to provide another possible
theoretical explanation for Chargaff's second parity rule, and to predict novel distributional aspects of
bacterial DNA sequences.
\end{abstract}

\bigskip

\bigskip
\hrule
\noindent
{\footnotesize\em This is a pre-copy-editing, author-produced preprint of an article accepted for publication in Biochemical and Biophysical Research Communications. The definitive publisher-authenticated version Marcelo Sobottka, Andrew G. Hart, A model capturing novel strand symmetries in bacterial DNA, Biochemical and Biophysical Research Communications, Volume 410, Issue 4, 15 July 2011, Pages 823-828, ISSN 0006-291X, is available online at: http://dx.doi.org/10.1016/j.bbrc.2011.06.072 or\break
http://www.sciencedirect.com/science/article/pii/S0006291X1101045X .}
\hrule
\bigskip


It is probable that the distributional structure of DNA sequences arises from
the accumulation of many successive stochastic events such as nucleotide
deletions, insertions, substitutions and elongations
(see \cite{Baldi2000,Bell97,Felsenstein81,Jukes69,Kimura80,Tavare89,Whittaker_et_al03}).
When studying a DNA sequence, the nucleotide frequencies and
dinucleotide frequencies are quantities often of interest to researchers.  The
nucleotide frequency $\pi_i$, for $i=A,C,G,T$, denotes the proportion of
type~$i$ bases in the sequence while the dinucleotide frequency
$P_{ij}$, for $i,j=A,C,G,T$, is the proportion of $2$-oligonucleotides of type $ij$ in the sequence. Given a two-stranded genome, let~$P$
and~$R$ be the matrices of dinucleotide frequencies for the primary
and complementary strands respectively according to the natural reading order on
each strand, and let~$\pi$
and~$\rho$ be the vectors of nucleotide frequencies for the primary
and complementary strands respectively.

Let us use $\alpha$ to denote the permutation which maps each
nucleotide to its complement ($\alpha(A)=T$, $\alpha(C)=G$, $\alpha(G)=C$ and
$\alpha(T)=A$).  We remark that $\pi_i=\rho_{\alpha(i)}$ and
$P_{ij}=R_{\alpha(j)\alpha(i)}$.

Chargaff's second parity rule says that the frequency of any short nucleotide sequence on a strand is approximately equal to the
frequency of its reverse complement on the same strand \cite{Baisnee,Chargaff,Rudner_et_al}.  For mononucleotides and dinucleotides,  this means that $\pi_i=\pi_{\alpha(i)}$ and
$\rho_i=\rho_{\alpha(i)}$, while $P_{ij}=P_{\alpha(j)\alpha(i)}$ and
$R_{ij}=R_{\alpha(j)\alpha(i)}$. Hence, for two-stranded
DNA, the first and second parity rules together are equivalent to $\pi=\rho$ and
$P=R$ element wise.
Recently, Mitchell and Bridge \cite{MitchellBridge} have found that, with the exception of the organellar DNA, all kinds of double stranded DNA genomes satisfy Chargaff's second parity rule.

We propose here a simple explanation for Chargaff's second parity rule for mononucleotides and dinucleotides, which is based only on the occurrence of random joinings of nucleotides in the sequence. Although such a model does not take account of evolution, coding sequences or other biological mechanisms (which should impose bias restrictions on the nucleotide frequencies in the DNA), it could explain the occurrence of Chargaff's second parity rule as a ``relic'' of the prebiotic DNA sequences which has been conserved throughout evolution (see
\cite{ZhangHuang08,ZhangHuang10,ZhangHuang10-2}). In particular, the model is consistent with Chargaff's second parity rule holding for many kinds of
double-stranded DNA, but not for single-stranded RNA/DNA (see \cite{MitchellBridge}).

We also use the model to predict various distributional aspects of
primitive DNA sequences and then, we compare distributional properties intrinsic to the model to statistical
estimates from 1049 bacterial DNA sequences.

We remark that, although the proposed model is based on Markovian processes, it doesn't imply that either primitive genomes or actual genomes are Markovian sequences of nucleotides. In fact, as we will see, the model here presented doesn't produce Markovian sequences (it produces sequences which are concatenations of Markovian sequences). Furthermore, actual genomes would be derived from many mutations of primitive DNA, which would introduce other non-Markovian features. On the other hand, although the existence of long-range correlations in non-coding portions of DNA sequences is well established (see \cite{Fukushima_et_al2000,Li92,Peng_et_al92,Yu_et_al2000})\sloppy, first order Markov
chains might well capture aspects of their nucleotide distributions (see \cite{Gao_et_al09}). This could explain why we are able to use the model here proposed to predict some distributional features of bacterial DNA.

\section*{Material and methods}

The 1049 complete genome sequences examined were downloaded from the GenBank
repository. 

We shall use $(\pi(n),P(n))$ and $(\rho(n),R(n))$ to denote the
pairs of mononucleotide and dinucleotide frequencies estimated for the
primary and complementary strands respectively of the $n^{\rm th}$ bacterium. Furthermore, we write $CG(n)$ for the $C+G$ content of the
$n^{\rm th}$ bacterium.

The bacteria analyzed satisfy
$\pi(n)\approx\rho(n)$ and $P(n)\approx R(n)$ as expected. In addition, we found that bacteria having similar $C+G$-contents seem to have similar matrices $P(n)$. Next, fixing $i,j\in\{A,C,G,T\}$, we can observe that the points $(CG(n),P_{ij}(n))$ for
$n=1,\ldots,1049$ appear to be systematically distributed around some curve
(see Fig. \ref{pi_x_P_theorical}).

The mononucleotide frequencies and dinucleotide frequencies were determined directly
from the sequences. We used the statistics and optimization toolboxes of
{\sc Matlab} to perform the data analysis on the bacterial DNA and estimate parameters of the model.\\

\noindent {\sc The model.}
We consider a model in which an actual genome is
obtained from uniformly distributed mutations on some primitive DNA sequence
which is constructed as follows: A new nucleotide is sporadically randomly selected as a
candidate for being  attached to one of the extremities of some strand of the
sequence duplex, and it joins the strand according to
some fixed probability. We make two assumptions about this construction
process: (A1) At each moment, each type of nucleotide has some probability of
being selected as a candidate for joining a strand (such probabilities
are supposed to be constant throughout the construction of each primitive DNA
sequence and could be interpreted as the availability of each nucleotide type in
the environment); (A2) The probability of a candidate nucleotide actually being
joined to the sequence depends on the type of the candidate nucleotide and the type
of the last nucleotide in the sequence (these probabilities are assumed to be
positive, constant and the same for all primitive DNA sequences, and could be
thought of as resulting from chemical and other physical properties of the
bases).

\begin{figure}[!h]
\centerline{\includegraphics[width=.4\linewidth=1.0]{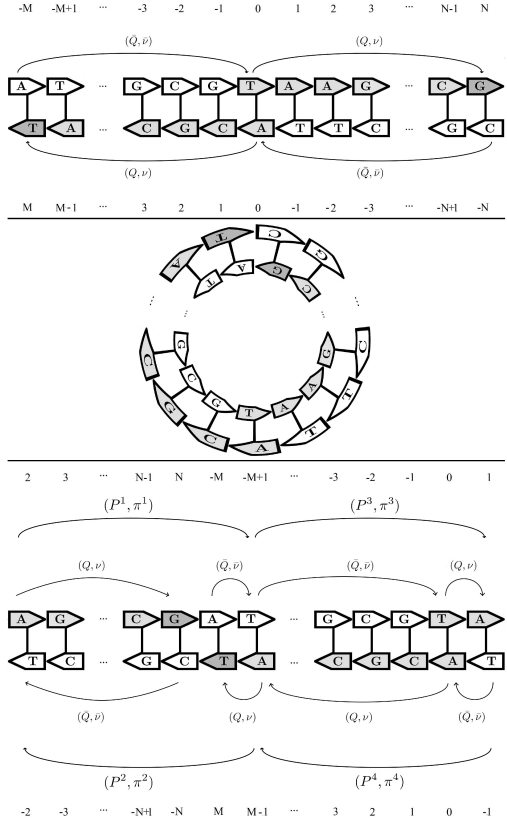}}
\caption{The top diagram shows a schematic representation of a
sequence produced by the model around an initial nucleotide pair at position
$0$. The illustration in the middle shows the circular DNA sequence obtained by
joining the extremities of the sequence. In the bottom diagram, a new sequence
is obtained by cutting the circular sequence at an arbitrary point. This final
sequence is a translation (cyclic shift) of the original one. If $M\approx N$
then the nucleotide distributions in the first (second) half of each strand are
very close to each other.}\label{DNAcircular}
\end{figure}

Let us represent a DNA duplex of length $L=M+N+1$ by the finite sequence
$\binom{x_\ell}{y_{-\ell}}_{-M\leq \ell\leq N}$, where
$x_\ell$ and $y_{-\ell}$ are nucleotides (which can be $A$, $C$, $G$ or $T$), and~$M$ and~$N$ are two positive integers (see Fig. \ref{DNAcircular}).  In
such a representation, $(x_\ell)_{-M\leq \ell\leq N}$ corresponds to the primary
strand while $(y_\ell)_{-N\leq \ell\leq M}$ corresponds to the complementary
strand.  The natural reading order for both strands is in the direction from~$-$
to~$+$, so that the strands are read in opposite directions within the duplex.
We want to describe the construction of each strand around an initial nucleotide pair
$\binom{x_0}{y_0}$.

Observe that $x_\ell = \alpha(y_{-\ell})$. Consequently,
knowledge of one strand is sufficient to reconstruct the entire duplex.  In a
similar way, knowledge of $(x_\ell)_{1\leq \ell\leq N}$ and $(y_\ell)_{1\leq \ell\leq M}$, together with
either $x_0$ or $y_0$, also suffices to specify the duplex (see Fig. \ref{DNAcircular}).

The model comprises two parts. Let the vector of probabilities
$\mu=(\mu_A,\mu_C,\mu_G,\mu_T)$ (assumed in~(A1)) be the relative abundance of
each nucleotide type in the environment.  Candidate nucleotide types are
selected at random according to these probabilities. Next, a primitive sequence of nucleotides is generated using a
matrix of probabilities $\aleph=(a_{ij})_{i,j=A,C,G,T}$ which
determines the suitability of candidate nucleotides for extending a strand.
Here, $a_{ij}$ represents the probability of a candidate nucleotide of type~$j$
being accepted and attached to a nucleotide of type~$i$ at the end of the
strand. Candidates that are
rejected remain in the environment with the possibility of being selected again
in the future.
	
Recall that the constructed sequences
$(y_\ell)_{-N\leq \ell\leq 0}$ and
$(x_\ell)_{-M\leq \ell\leq 0}$ do not
necessarily have the same mononucleotides and dinucleotide frequencies as $(x_\ell)_{0\leq
\ell\leq N}$ and $(y_\ell)_{0\leq \ell\leq M}$. In fact, the observable mononucleotide and dinucleotide frequencies of $(x_\ell)_{0\leq \ell\leq N}$ and $(y_\ell)_{0\leq \ell\leq M}$ are
given by the vector $\nu=(\nu_A,\nu_C,\nu_G,\nu_T)$ and the matrix $Q=(Q_{ij})_{i,j=A,C,G,T}$,
while the observable
mononucleotide and dinucleotide frequencies of $(y_\ell)_{-N\leq \ell\leq 0}$ and $(x_\ell)_{-M\leq \ell\leq
0}$ are determined by the vector $\bar{\nu}=(\bar{\nu}_A,\bar{\nu}_C,\bar{\nu}_G,\bar{\nu}_T)$ and the matrix $\barQ=(\barQ_{ij})_{i,j=A,C,G,T}$, where $\barnu_i=\nu_{\alpha(i)}$ and $\barQ_{ij}=Q_{\alpha(j),\alpha(i)}$ (see Appendix A for mathematical details).

Notice that each of the sequences $(x_\ell)_{-M\leq\ell\leq N}$ and $(y_\ell)_{-
N\leq\ell\leq M}$ produced by the model is a concatenation of the origin with
two Markovian sequences, one of length~$N$ and the other of length~$M$. Furthermore,
as~$L$ increases, $t:=N/L$ tends to the proportion of the primary strand
whose mononucleotide frequency is $\nu$ and dinucleotide frequency is~$Q$ while $1-t$ approaches the proportion of the primary strand
whose mononucleotide frequency is $\barnu$ and dinucleotide frequency is~$\barQ$. These proportions are $1-t$ and~$t$, respectively, in
the complementary strand. Thus, if~$L$ is large, then the mononucleotide and dinucleotide frequencies estimated for $(x_\ell)_{-M\leq\ell\leq N}$ and $(y_\ell)_{-
N\leq\ell\leq M}$ are respectively approximated by
\begin{equation}
\label{P_pi_R_rho}
\begin{array}{lcl}
\displaystyle P_{ij}
&=&\displaystyle tQ_{ij}+(1-t)Q_{\alpha(j)\alpha(i)}, \\\\
\displaystyle \pi_i
&=&\displaystyle t\nu_i+(1-t)\nu_{\alpha(i)}\\\\\\
\text{and}\\\\\\
\displaystyle R_{ij}
&=&\displaystyle (1-t)Q_{ij}+tQ_{\alpha(j)\alpha(i)}, \\\\
\displaystyle \rho_i
&=&\displaystyle (1-t)\nu_i+t\nu_{\alpha(i)}.
\end{array}
\end{equation}

\section*{Results}

\noindent {\sc Deriving Chargaff's second parity rule from the model.}
Simulations of the model for many distinct vectors $\mu$, matrices $\aleph$ and values of $t$, show that, if the entries in the rows of the matrix $\aleph$ are very different from each other and the value of $t$ is far from 0.5, then the sequences produced by the model in general don't satisfy Chargaff's second parity rule for mononucleotides and dinucleotides.

    For the specific case where $t=0.5$, then equations~\eqref{P_pi_R_rho}
	imply $\pi=\rho$ and $P=R$, independently of the matrix $\aleph$.  This could possibly
	explain why many double-stranded genome sequences comply with Chargaff's second parity
	rule for mononucleotides and dinucleotides. In fact, if a primitive DNA duplex were constructed according to the
	model without one strand being favored over the other (that is, $N$
	significantly smaller or larger than~$M$), then in general we would have
	$t\approx 0.5$. Furthermore, if a genome resulted from relatively few
	mutations  distributed uniformly throughout some primitive DNA sequence, then
	we would expect to see the original distributional structure preserved along
	large segments of the sequence. In particular, although elongations (repetitions of
	parts of the sequence separated by arbitrarily long distances) would generate
    long-range correlations, the resulting
	sequence would have similar distributions in both the original and the new part,
	while only a small variation would appear in the position where the new part
	was concatenated with the original one.\\

\noindent {\sc Predicting regions of the genomes with similar mononucleotide and dinucleotide frequencies.}
The occurrence of $t\approx 0.5$ with mutations distributed uniformly throughout
the sequence would also imply that the mononucleotide and dinucleotide frequencies for the first
(second) half of each strand are closer to each other than the mononucleotide and dinucleotide frequencies
over any other half (see Fig. \ref{DNAcircular}). We remark that the fact that bacterial DNA is generally circular
would not affect such a property, since each ``linearized'' DNA sequence would be a translation of some sequence produced by
the model (also to see Fig. \ref{DNAcircular}).

We have examined
this property in bacterial genomes. Each strand of the 1049 bacterial DNA
sequences was partitioned into two parts of equal length. For the $n^{\rm th}$
bacterium, we use $(\pi^1(n),P^1(n))$ and $(\pi^4(n),P^4(n))$ to denote the
mononucleotide and dinucleotide frequencies computed from the first half of the primary and complementary
strands, respectively, and we let $(\pi^3(n),P^3(n))$ and $(\pi^2(n),P^2(n))$ be
the mononucleotide and dinucleotide frequencies computed from the second half of the primary and
complementary strands, respectively. Then, we compared the mononucleotide and dinucleotide frequencies of each DNA segment in each one of the bacterial genomes, estimating the coefficients of the linear regressions which best fit the data (see Table \ref{Four_parts_test}).


\begin{table}
\caption{The coefficients estimated from the 1049 bacterial genomes for the linear regressions which best fit the relationship of mononucleotide and dinucleotide frequencies between each segment of DNA. For example, the relationship between the AC frequency $P^{1}_{AC}$ on segment 1 and the AC frequency $P^{4}_{AC}$ on segment 4 is fitted by the relation $P^{4}_{AC}=0.974P^{1}_{AC}+0.001$ with $R^2=0.955$ (the entry in row `AC' and column `1-4').\label{Four_parts_test}}

\tiny
{\begin{tabular}{|p{0.3cm}|cccccc|}
\hline &  & & & & &\\
 &   1-2       &  1-3       & {\bf 1-4}        & {\bf 2-3}   & 2-4    & 3-4\\
 &   $\overline{slope\  intercept  \ R^2}$       &  $\overline{slope\  intercept \  R^2}$       & $\overline{\mathbf{slope\  intercept \  R^2}}$       & $\overline{\mathbf{slope\  intercept \  R^2}}$   & $\overline{slope\  intercept \  R^2}$    & $\overline{slope\  intercept \  R^2}$\\
\hline
A         &  0.876  \ \ 0.031 \ \  0.930  &  0.873 \ \   0.032  \ \  0.929  & {\bf 1.002  \ \  0.000  \ \  0.994 } & {\bf 0.993  \ \  0.001  \ \  0.993 } &  1.066  \ \ -0.016  \ \  0.930  &  1.065  \ \ -0.015  \ \  0.922  \\
C         &  0.894  \ \ 0.043 \ \  0.921  &  0.894 \ \   0.043  \ \  0.924  & {\bf 0.997  \ \  0.001  \ \  0.995 } & {\bf 0.996  \ \  0.001  \ \  0.993 } &  1.033  \ \ -0.026  \ \  0.926  &  1.033  \ \ -0.026  \ \  0.924  \\
G         &  1.031  \ \-0.026 \ \  0.921  &  1.033 \ \  -0.026  \ \  0.926  & {\bf 0.996  \ \  0.001  \ \  0.993 } & {\bf 0.997  \ \  0.001  \ \  0.995 } &  0.894  \ \  0.043  \ \  0.924  &  0.895  \ \  0.043  \ \  0.924  \\
T         &  1.062  \ \-0.015 \ \  0.930  &  1.066 \ \  -0.016  \ \  0.930  & {\bf 0.993  \ \  0.001  \ \  0.993 } & {\bf 1.002  \ \  0.000  \ \  0.994 } &  0.873  \ \  0.032  \ \  0.929  &  0.866  \ \  0.033  \ \  0.922  \\
AA        &  0.885  \ \ 0.008 \ \  0.928  &  0.882 \ \   0.008  \ \  0.924  & {\bf 1.003  \ \  0.000  \ \  0.995 } & {\bf 0.994  \ \  0.000  \ \  0.992 } &  1.054  \ \ -0.003  \ \  0.927  &  1.051  \ \ -0.003  \ \  0.918  \\
AC        &  0.648  \ \ 0.021 \ \  0.339  &  0.662 \ \   0.021  \ \  0.359  & {\bf 0.974  \ \  0.001  \ \  0.955 } & {\bf 0.972  \ \  0.002  \ \  0.960 } &  0.537  \ \  0.020  \ \  0.360  &  0.531  \ \  0.020  \ \  0.347  \\
AG        &  0.482  \ \ 0.025 \ \  0.275  &  0.494 \ \   0.024  \ \  0.292  & {\bf 0.988  \ \  0.001  \ \  0.957 } & {\bf 0.970  \ \  0.002  \ \  0.951 } &  0.586  \ \  0.026  \ \  0.284  &  0.567  \ \  0.027  \ \  0.263  \\
AT        &  1.000  \ \ 0.000 \ \  1.000  &  0.997 \ \   0.000  \ \  0.997  & {\bf 0.997  \ \  0.000  \ \  0.997 } & {\bf 0.997  \ \  0.000  \ \  0.997 } &  0.997  \ \  0.000  \ \  0.997  &  1.000  \ \  0.000  \ \  1.000  \\
CA        &  0.856  \ \ 0.014 \ \  0.459  &  0.885 \ \   0.013  \ \  0.492  & {\bf 0.964  \ \  0.002  \ \  0.940 } & {\bf 0.977  \ \  0.002  \ \  0.955 } &  0.555  \ \  0.023  \ \  0.497  &  0.543  \ \  0.024  \ \  0.474  \\
CC        &  0.935  \ \ 0.011 \ \  0.916  &  0.939 \ \   0.011  \ \  0.920  & {\bf 0.995  \ \  0.000  \ \  0.992 } & {\bf 0.998  \ \  0.000  \ \  0.991 } &  0.980  \ \ -0.006  \ \  0.920  &  0.977  \ \ -0.006  \ \  0.918  \\
CG        &  1.000  \ \ 0.000 \ \  1.000  &  0.998 \ \   0.000  \ \  0.998  & {\bf 0.998  \ \  0.000  \ \  0.998 } & {\bf 0.998  \ \  0.000  \ \  0.998 } &  0.998  \ \  0.000  \ \  0.998  &  1.000  \ \  0.000  \ \  1.000  \\
CT        &  0.571  \ \ 0.027 \ \  0.275  &  0.586 \ \   0.026  \ \  0.284  & {\bf 0.970  \ \  0.002  \ \  0.951 } & {\bf 0.988  \ \  0.001  \ \  0.957 } &  0.494  \ \  0.024  \ \  0.292  &  0.464  \ \  0.026  \ \  0.263  \\
GA        &  0.662  \ \ 0.016 \ \  0.272  &  0.702 \ \   0.013  \ \  0.306  & {\bf 0.963  \ \  0.003  \ \  0.928 } & {\bf 0.979  \ \  0.001  \ \  0.959 } &  0.425  \ \  0.039  \ \  0.292  &  0.415  \ \  0.039  \ \  0.279  \\
GC        &  1.000  \ \ 0.000 \ \  1.000  &  0.995 \ \   0.000  \ \  0.998  & {\bf 0.995  \ \  0.000  \ \  0.998 } & {\bf 0.995  \ \  0.000  \ \  0.998 } &  0.995  \ \  0.000  \ \  0.998  &  1.000  \ \  0.000  \ \  1.000  \\
GG        &  0.980  \ \-0.006 \ \  0.916  &  0.980 \ \  -0.006  \ \  0.920  & {\bf 0.998  \ \  0.000  \ \  0.991 } & {\bf 0.995  \ \  0.000  \ \  0.992 } &  0.939  \ \  0.011  \ \  0.920  &  0.940  \ \  0.011  \ \  0.918  \\
GT        &  0.522  \ \ 0.020 \ \  0.339  &  0.537 \ \   0.020  \ \  0.360  & {\bf 0.972  \ \  0.002  \ \  0.960 } & {\bf 0.974  \ \  0.001  \ \  0.955 } &  0.662  \ \  0.021  \ \  0.359  &  0.653  \ \  0.021  \ \  0.347  \\
TA        &  1.000  \ \ 0.000 \ \  1.000  &  0.999 \ \   0.000  \ \  0.998  & {\bf 0.999  \ \  0.000  \ \  0.998 } & {\bf 0.999  \ \  0.000  \ \  0.998 } &  0.999  \ \  0.000  \ \  0.998  &  1.000  \ \  0.000  \ \  1.000  \\
TC        &  0.410  \ \ 0.040 \ \  0.272  &  0.425 \ \   0.039  \ \  0.292  & {\bf 0.979  \ \  0.001  \ \  0.959 } & {\bf 0.963  \ \  0.003  \ \  0.928 } &  0.702  \ \  0.013  \ \  0.306  &  0.671  \ \  0.015  \ \  0.279  \\
TG        &  0.536  \ \ 0.024 \ \  0.459  &  0.555 \ \   0.023  \ \  0.497  & {\bf 0.977  \ \  0.002  \ \  0.955 } & {\bf 0.964  \ \  0.002  \ \  0.940 } &  0.885  \ \  0.013  \ \  0.492  &  0.874  \ \  0.013  \ \  0.474  \\
TT        &  1.049  \ \-0.003 \ \  0.928  &  1.054 \ \  -0.003  \ \  0.929  & {\bf 0.994  \ \  0.000  \ \  0.992 } & {\bf 1.003  \ \  0.000  \ \  0.995 } &  0.882  \ \  0.008  \ \  0.924  &  0.874  \ \  0.009  \ \  0.918  \\
\hline
\end{tabular}\\\\
}
\normalsize

\end{table}

Note that with the exception of the dinucleotides $AT$, $CG$, $GC$ and $TA$, the genomes examined satisfy the property which was predicted by the model. The failure of this property for the dinucleotides $AT$, $CG$, $GC$ and $TA$ has a simple explanation: Due to Chargaff's
first parity rule a dinucleotide of one of these types corresponds exactly to a dinucleotide of the same type on the complementary strand. Thus these dinucleotides occur with exactly the same frequencies on segments 1 and 2 of the DNA duplex and also occur with exactly the same frequencies on segments 3 and 4 of the DNA duplex.\\


\noindent {\sc Estimating parameters and finding properties of $\aleph$.}
From assumption (A1), if a nucleotide is proposed as a candidate for joining the end of one
of the strands, then its complement is automatically a candidate for joining
the beginning of the other strand. Therefore, the probability of a nucleotide
being selected as a candidate for joining one strand is the same as the probability of its
complement being selected for the other strand, that is,\\

\noindent{\bf ($H1$)} $\mu_A=\mu_T$ and $\mu_C=\mu_G$.  In other
words, $\mu$ takes the form $\mu(m)=(m,0.5-m,0.5-m,m)$, where $0\leq m \leq
0.5$.\\

Now, since the matrix~$\aleph$ is assumed to be the same for all genomes, any
sequence produced by the model is a realization of a Markov chain belonging to a
family of Markov chains parameterized by~$m$ and~$t$, where  $0< m< 0.5$ and
$0\leq t\leq 1$.

Further, from (A2) we have that the probability of a candidate nucleotide of
type~$j$ being accepted to follow a nucleotide of type~$i$ at the end of one
strand is equal to the probability of a nucleotide of type $\alpha(i)$ preceding
a nucleotide of type $\alpha(j)$ at the beginning of the other strand, that is,
the matrix $\aleph$ has the form\\

\noindent{\bf ($H2$)} $a_{ij}=a_{\alpha(j)\alpha(i)}$.\\

We use the {\em lsqcurvefit} function of the optimization toolbox in
{\sc Matlab} to construct estimators for the matrices $\aleph(n)$, vectors~$\mu(n)$  and values $t(n)$. The {\em
lsqcurvefit} function solves (in the least-squares sense) the nonlinear
problem of determining the parameters for which the right side of equation
\eqref{P_pi_R_rho} most closely approximates~$P(n)$. Firstly, we carried
out free estimation without the {\em a priori} assumption of the properties ($H1$) and
($H2$).

The solution to the optimization problem is sensitive to the value of~$t(n)$
chosen to initialize the optimization algorithm. Since previous tests suggest
that neither strand is significantly favored over the other during the
construction of the sequence, we used $t(n)=0.5$ as the initial value.
Furthermore, the optimization process only permits the estimation of the
probability of a nucleotide of type $j$ joining a nucleotide of type $i$
relative to the probability of a nucleotide of type~$k$ joining a nucleotide of
type~$i$. In order that $\aleph(n)$ be comparable in a meaningful way between
different numerical experiments, it was scaled after the optimization
procedure to have the sum of all of its elements equal to ten.

We found that the vectors $\mu(n)$ estimated for each matrix $P(n)$ generally satisfy the property ($H1$). The mean absolute difference between $\mu_i(n)$ and $\mu_{\alpha(i)}(n)$ was found to be 0.0254, while the median absolute difference was 0.0198. Furthermore, we computed the average
$$\bar{\aleph}=\left(\begin{matrix}
 \displaystyle   0.7217  & \displaystyle 0.5236  & \displaystyle 0.5908  & \displaystyle 0.6672\\
 \displaystyle   0.6815  & \displaystyle 0.6055  & \displaystyle 0.6138  & \displaystyle 0.5986\\
 \displaystyle   0.6507  & \displaystyle 0.7187  & \displaystyle 0.6035  & \displaystyle 0.5304\\
 \displaystyle   0.4548  & \displaystyle 0.6420  & \displaystyle 0.6758  & \displaystyle 0.7213\\
    \end{matrix}\right)$$
of all 1049 estimated matrices $\aleph(n)$. To assess whether or not we can interpret $\bar{\aleph}$ as
having Property ($H2$), we estimated~$\aleph(n)$, $\mu(n)$ and
$t(n)$ again, this time imposing the restrictions stipulated by ($H1$)
and ($H2$). In this case we obtained the following average:
$$\bar{\bar{\aleph}}=\left(\begin{matrix}
 \displaystyle   0.7515 & \displaystyle   0.4807 & \displaystyle   0.5583 & \displaystyle   0.6785\\
 \displaystyle   0.6942 & \displaystyle   0.5584 & \displaystyle   0.6141 & \displaystyle   0.5583\\
 \displaystyle   0.6722 & \displaystyle   0.7407 & \displaystyle   0.5584 & \displaystyle   0.4807\\
 \displaystyle   0.5361 & \displaystyle   0.6722 & \displaystyle   0.6942 & \displaystyle   0.7515\\
 \end{matrix}\right).$$

\begin{figure}[!h]
\centerline{\includegraphics[width=.7\linewidth=1.0]{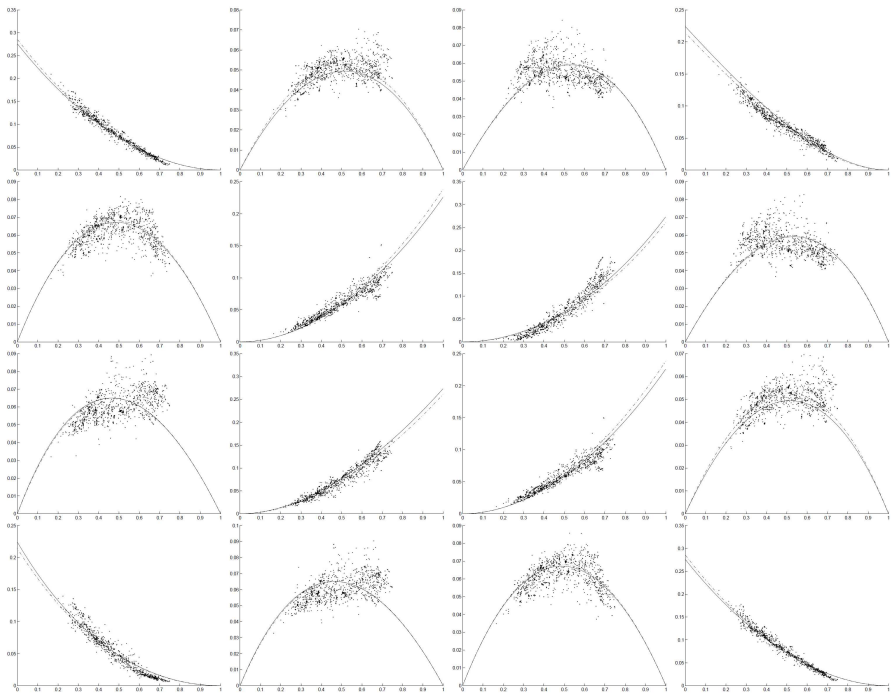}}
\caption{The graph in the $i^{\rm th}$ row and $j^{\rm th}$ column
plots the points $(CG(n),P_{ij}(n))$ for the 1049 bacteria. Then,
fixing $t=0.5$ and using the matrices~$\bar{\aleph}$ and~$\bar{\bar{\aleph}}$,
we computed the matrices $\bar{P}(m)$ and $\bar{\bar{P}}(m)$ respectively for many
distinct values of $\mu(m)=(m,0.5-m,0.5-m,m)$. The points
$(\overline{CG}(m),\bar{P}_{ij}(m))$ (dashed line) and
$(\overline{\overline{CG}}(m),\bar{\bar{P}}_{ij}(m))$ (solid line) have been plotted on the
appropriate graph in the $i^{\rm th}$ row and $j^{\rm th}$ column.}\label{pi_x_P_theorical}
\end{figure}

We can use
$\bar{\aleph}$ and $\bar{\bar{\aleph}}$ with distinct values of $m\in(0,\ 0.5)$
to produce mononucleotide and dinucleotide frequencies $(\bar{\pi}(m),\bar{P}(m))$ and
$(\bar{\bar{\pi}}(m),\bar{\bar{P}}(m))$, respectively, according to equation
\eqref{P_pi_R_rho} with $t=0.5$. Then, plotting the points
$(\overline{CG}(m),\bar{P}_{ij}(m))$ and
$(\overline{\overline{CG}}(m),\bar{\bar{P}}_{ij}(m))$, we obtained curves that were
very close to each other.  Furthermore, the majority of the points $(CG(n),P_{ij}(n))$ are
distributed around those curves (see
Fig. \ref{pi_x_P_theorical}). This suggests that the construction
process defined by $\bar{\aleph}$ is close to one defined using a matrix which
satisfies ($H2$).

\section*{Discussion}

There are two alternative approaches to explaining Chargaff's
second parity rule. Indeed, it can either be supposed to arise
from evolutionary convergence caused by mutation and selection
(see \cite{AlbrechtBuehler2006,AlbrechtBuehler2007,Fickett_et_al1992,Lobry1995,LobryLobry1999,Sueoka1995}) or it can be supposed to be a
characteristic of the primordial genome (a ``relic'')
\cite{ZhangHuang08,ZhangHuang10,ZhangHuang10-2}).

In this work we have proposed a stochastic model which is consistent with Chargaff's second parity rule
being a characteristic of primordial double stranded DNA
genomes. Besides providing a simple explanation for Chargaff's
second parity rule, the model
predicts other distributional properties of bacterial DNA
sequences. In particular, it can be used to predict regions of
DNA sequences that are distributionally close to each
other.

We remark that Chargaff's second parity rule, together with the
property that the nucleotide distributions for the first
(second) half of each strand are closer to each other than to
the distribution over any other half, can be derived from any
stochastic construction around an initial nucleotide pair
analogously to the way we have described, that is, neither
phenomenon requires such construction to be Markovian in
nature. Markovianness was assumed here merely as a first step towards investigation.  In fact, if $(x_\ell)_{0\leq \ell\leq N}$ and
$(y_\ell)_{0\leq \ell\leq M}$ were generated using the same
non-Markovian law, we would observe the same properties. On the
other hand, the fact that we have found good agreement with
Properties ($H1$) and ($H2$) in the
estimation of~$\aleph$ and~$\mu$ suggests that the Markovian
construction of primitive DNA sequences succeeds
in capturing the gross structure at the level of dinucleotides.

We could also extend the proposed model to the case where $\mu$
varies during the construction process. In such a case, we
could derive an equation like \eqref{P_pi_R_rho}, and use it to
explain Chargaff's second parity rule. However, if $\mu$ varies
with time, then we should not observe $P^1(n)$ closer to
$P^4(n)$ than to $P^2(n)$ or $P^3(n)$, even when $t=0.5$. This, together with the
fact that our estimation returned a $\bar{\aleph}$ which
provides a reasonable approximation to the mononucleotide and
dinucleotide frequencies $(\pi(n),P(n))$, seems to
indicate that~$\mu$ does not vary much during the construction
process.

We recall that this work has a primarily speculative purpose and further work should address the nucleotide distribution phenomena indicated by this model. In particular, rigorous (biological) criteria to select the tested genomes should be taken in to account to avoid biased estimations when we use the model to predict regions of the DNA with the same nucleotide distribution. Finally, the properties of the estimators used in the statistical
analyses merit study and other estimators could be considered. In
addition, a simulation study could assess the sensitivity to mutation
rate on the ability to recover properties of the original sequences.

\section*{Appendix A. The mathematics of the model}

The construction of $(x_\ell)_{0\leq \ell\leq N}$ and $(y_\ell)_{0\leq
\ell\leq M}$ proceeds according to (A1) and (A2) as follows: Suppose the primary
strand has been extended from the origin~$0$ up to the point~$\ell$, which is a
nucleotide of type~$i$. With probability $\mu_j$, a nucleotide of type~$j$ is
selected as a candidate to join the strand. The probability of the candidate
being accepted as the next nucleotide in the sequence, given that it is of
type~$j$, is $a_{ij}$. This process corresponds to the coupling of an urn
process characterized by~$\mu$ with a Markov chain whose transition matrix is
obtained by normalizing the rows of~$\aleph$.

Thus, the observable conditional dinucleotide frequencies of $(x_\ell)_{0\leq \ell\leq N}$ and $(y_\ell)_{0\leq \ell\leq M}$ are given by the matrix $W=(W_{ij})_{i,j=A,C,G,T}$

\begin{equation}
\label{processQ}
W_{ij}=\frac{a_{ij}\mu_j}{\sum_k a_{ik}\mu_k}.
\end{equation}

The stationary distribution of $W$ is the vector $\nu=(\nu_A,\nu_C,\nu_G,\nu_T)$, which corresponds to the mononucleotide frequencies of $(x_\ell)_{0\leq \ell\leq N}$ and $(y_\ell)_{0\leq
\ell\leq M}$ if these sequences are sufficiently long. Therefore the dinucleotide frequencies are obtained from the expression $Q_{ij}=\nu_i W_{ij}$.

%

\section*{Acknowledgments}
This work was supported by the Basal CONICYT program PFB 03
hosted by the Center for Mathematical Modeling (CMM) at the University of Chile.
M. Sobottka was supported by CNPq-Brazil grant 304457/2009-4 and by FUNPESQUISA/UFSC 2009.0138. The
authors thank the Laboratory of Bioinformatics and Mathematics of the Genome in
the CMM for assisting in the acquisition of data and providing advice.

\bibliographystyle{plain}
\bibliography{Chargaff_JMB}   


\end{document}